\documentclass[11pt,twoside]{article}
\usepackage{newpasp}
\def\edcomment#1{\iffalse\marginpar{\raggedright\sl#1\/}\else\relax\fi}
\reversemarginpar

\begin{document}
\title{Particles and Fields in Radio Galaxies: A Summary}
\author{R. D. Blandford}
\affil{130-33 Caltech, Pasadena, CA 91125, USA}
\begin{abstract}
A summary is presented of a meeting on Particles and Fields 
in Radio Galaxies held in Oxford in August 2000. Recent detailed
studies of radio maps and the first X-ray images from Chandra X-ray
Observatory, together with new capabilities in simulating 
hydromagnetic flows, are transforming our understanding of how jets
are formed by accretion disks orbiting massive black holes, how these
jets dissipate and how they inflate the giant radio-emitting lobes
by which they were first identified. As they can be imaged 
in such detail, extragalactic radio sources remain central
to the study of jet outflows in general although there is a 
strong convergence with corresponding studies of Galactic superluminal
sources, gamma ray bursts and young stellar objects.
\end{abstract}
\section{Introduction}
This meeting, which was planned as a specialist discussion on the details
of emission mechanisms in extragalactic radio sources, was held at a 
propitious time. The first X-ray images of jets and their associated
lobes from Chandra X-ray Observatory had just been 
released and, just as was the case when linked and 
very long baseline radio interferometry became 
possible and when we learned of the prodigious
$\gamma$-ray powers of blazars, a whole new chapter in the 
eighty year old study of jets is being started. The study of extragalactic
radio sources has also changing in a quite different sense because
the jet phenomenon is turning out to be far more common in its astronomical
expression than extragalactic astronomers ever expected, as it is 
now known to be 
a common accompaniment to young stellar objects
and an occasional by-product of accretion onto compact stars.
 
Nevertheless, the main topics of discussion at this meeting
mostly centred around the 
radio study of extragalactic jets and I have chosen to organise 
this somewhat impressionistic summary using a series of general questions
that have been around for a while and where, although we don't yet have 
answers, we are making progress. I will use the convention of referring
to contributors using [ ] and defer to these authors for original references.
\section{Jets}
\subsection{Why do Accreting Sources Form Jets?}
If I follow the logic of the theorist as opposed to that of the observer,
this is the first question to ask and it is one that was implicit
in many of the theoretical papers. A somewhat 
non-standard answer is that when sources accrete, 
the gas that they collect almost
certainly possesses far more specific angular momentum than the accreting
object and almost all of this must be off-loaded. 
Now, ever since the first discussion
of disks in binary and proto- stars, we have been acutely aware that there
has to be a strong torque that accomplishes this task and we now know that
this torque is almost always hydromagnetic in character. The gas forms a 
disk flow and the differential rotation drives the exponential growth of 
magnetic field on a dynamical time, until it, too, is dynamically important.
As matter flows in,  angular momentum flows out; at least that is what
has been conventionally assumed. However, it need not all flow out.
Disks have a lot of surface area and
there {\it may} be an appreciable divergence in the angular momentum 
flux due to surface magnetic torque. After all,
the sun, a very slow rotator by the standards of an accretion disk, 
has managed to lose a significant fraction of its initial angular momentum
though the solar wind in this manner. Although we do not understand
how much of the disk angular momentum escapes through this 
channel, a jet, or more
generally, a bipolar outflow, may be simply 
Nature's way of removing unwanted
angular momentum. 

The torque, $G$, must also do work at a rate $G\Omega$, where $\Omega$
is the angular velocity. Not only is angular momentum transported outward
through the flow, so is energy. In fact, if the gas is gravitationally
bound, there {\it must} be a divergence of this flow of energy. 
In a traditional
accretion disk, this is the source of the radiant energy and this explains
why the energy release in 
simple models is three times the locally liberated binding
energy. However, not all flows will radiate efficiently and if they don't
we have a second possible answer to the question. Jets (or,
more generally, outflows)
are the next best way to remove the energy from the disk
when radiation is not up to the job.

Disks are not the only possible sources of 
power and the central object can contribute
significantly. In the case of massive black holes in AGN, the rotational 
energy of the hole -- surely ample to re-supply the most profligate
of radio lobes -- can be extracted with modest efficiency
using electromagnetic field. This can be directly released 
in a jet core in the form of Poynting
flux [Meier, van Putten, Begelman] or, alternatively,
transmitted to the inner disk. In 
the former case the energy released is likely to have a low
baryon fraction and be in an ideal state to power ultrarelativistic jets.
Even if the central object is very slowly rotating, as appears to 
be true of protostars and many neutron stars, there is still as much 
orbital kinetic energy to tap from the innermost disk orbit as
has been released in the form of binding energy up to this point. 

Given these different options, the question becomes
``How much energy flows along each of these channels under
different circumstances?''.
We know that hydromagnetic jets do not always remove all the 
angular momentum and energy, despite the
fact that there is no dynamical objection to this 
happening. After all disks are prodigiously luminous as quasars
and binary X-ray sources. Perhaps, all disks lose some angular momentum
and energy  
in this manner but that it is only those that {\it need} to get rid of 
{\it energy}
-- specifically those whose central masses are so large as to render
the mass supply rate sub-critical and thermal electron heating 
ineffectual--
that produce the most efficient disk outflows. In addition,
perhaps it is only those 
sources where the central hole (or star) spins fast that create
high speed flows and that both factors are neccesary 
to form high power ultra-fast jets.  
\subsection{How do Accreting Sources Make Jets?}
Following this theoretical tack, there has been much progress,
some of it reported at the meeting [Begelman, Meier, Trussoni, Hughes]
towards developing an understanding
of the global behaviour of disk magnetic fields. The issue is really that
the sort of investigations that are possible analytically are usually 
artifically limited to assumptions like high symmetry and self-similarity
and these can be quite misleading. 
Furthermore, even if we acknowledge that some simple
solution of the equations of MHD is formally unstable, this may be 
misleading because nonlinear saturation can limit its importance.
Magnetic field is famously slippery and any time we have been able to 
observe it in action -- in the heliosphere or the plasma laboratory -- it has 
demonstrated far more imagination than us. For this reason,
the widepread use of three dimensional MHD codes (in some cases,
general relativistic) is a great advance. True, 
they have to leave out a lot of the physics like reconnection, radiation,
Alfv\'en wave damping and so on, but still they allow us to perform quite 
novel experiments which sharpen our intuition as to what is possible.

If one accepts, as most now do, that jets and outflows 
are fundamentally relativistic and hydromagnetic,
there are still several options. (An important
exception may be BALQ and even Seyfert 
outflows which are sub-relativistic and appear to be 
associated with objects radiating near to the Eddington
limit [Wilson].) Jets can be powered by 
the disk or the hole, or both. Jets may be launched
centrifugally or through pressure, magnetic, radiative or gas. They may 
be collimated by an organised toroidal field, as is found in the solar wind,
or by a much more disorganised assembly of twisted field loops. All of these
models have had enthusiastic proponents and simulations are sharpening
the debate. As observations of nearby jets reach down to smaller
radii, $\sim100m$ in the case of M87 [Sparks].

New phenomena, like magnetic switching [Meier] have been 
discovered. Perhaps we are learning that hydromagnetic jets are intrinsically
episodic and that we have been leaving out a key ingredient 
when we have taken the easy way out, at least in analytical
modelling, by seeking stationary solutions to the flow equations. Surely
the observations -- of QPOs in GRS1915+105, of superluminal expansion
in compact quasars and of Herbig Haro objects -- support this 
interpretation. 

A key concern is ``Can a disk maintain a large scale, poloidal
magnetic field or will this field migrate outward under resistive diffusion
faster than it is convected inward under flux-freezing?''. Another fundamental
physics concern is what happens to the energy that is dissipated in a disk 
at a rate $G\nabla\Omega$? Does it all feed a hydromagnetic wave turbulent 
cascade being damped at small length scale through wave-particle interaction?
Alternatively, does it get converted into heat through magnetic reconnection?
Should we introduce a second parameter $\alpha'$, that takes account of the 
dissipation? These considerations are also directly
relevant to the jet flows themselves where linear velocity gradients
take the place of differential rotation. 

This leads me to the admission, that every theorist who has worked in this 
field must make, which is that the study of jets
is an empirical, observational business
at least when we try to go much beyond fundamental conservation laws. (Not all
astrophysics is likely this!)  There is a level of detail in the radio 
maps which we can try to describe but which we would
not have predicted from first principles. If this proposition needs
defense, and in a meeting attended mainly by observers I doubt
if it does, then the recent Chandra images of the X-ray nebulae
surrounding the Crab and Vela pulsars should surely make the point.
In both cases, isolated, spinning, magnetised neutron stars are,
quite unexpectedly,  able to create
apparently well-focused outflowing jets
presumably without the aid of a disk. They also show evidence for large
scale, transverse shocks and equatorial ring structures. 
Magnetic fields are again
the best way to try to explain the observations. (The equatorial ring-like
structure might be attributable to a current sheet formed in an outflowing 
wind, analogous to what happens with the solar wind.) However, we do not
yet know how this happens.

As we cannot answer any of these basic MHD questions
through theory alone and so we must turn to the laboratory. 
The sophistication of the 
diagnostics of the terrestrial magnetosphere and, especially, the 
solar corona, through the YOHKOH, SOHO and most recently TRACE spacecraft,
is truly impressive. There is a wealth of data, publicly available, that
is large ignored that ought to contain answers to some of these questions.
Let me give two examples. It appears that most of the reconnection in the 
solar flares takes place on the small scale, in nanoflares, rather than 
in the giant flares that disrupt communications and so on
[Beckert]. If the same is true
of an accretion disk corona then this has serious implications for our view 
of how a disk corona 
is heated. If we look at the solar wind, we find that
it emerges from a gas whose temperature does not exceed 2 million degrees
and yet it can achieve outflows speeds in excess of 800 km s$^{-1}$. It is 
launched neither thermally or centrifugally. Perhaps it is sent
on its way by acquiring the momentum of outflowing hydromagnetic waves.
Observations are starting to settle these matters in some generality.
\subsection{How do we ``see'' Jets?}
A jet is an expanding outflow and as such the natural
synchroton emissivity should decline rapidly with radius. The rate is 
model-dependent, of course, but is generally 
much faster than observed [Giovannini]. For this
reason, it has been assumed that there is {\it in situ}
particle acceleration along the jet. This fits in well with the notion
that jets are naturally episodic, perhaps on account of instabilities
that develop near their sources. These instabilites will naturally 
develop into internal shocks as the flows are hypersonic. They will 
also be responsible for an overall deceleration of the flow as
deduced on spectral grounds [Laing]. In addition,
the termination shocks are associated with hot spots [Brunetti, Hardcastle].

Now non-relativistic shocks appear to be responsible
for transmitting a power law distribution of relativistic ions
and electrons in supernova remanants through diffusive Fermi acceleration.
According to the basic theory of this process, the efficiency is high
and the back pressure associated with the reflected ions
may provide a negative feedback on the process. However, we still
do not have a good theory of the magnetic turbulence that can explain
how particles are back-scattered [Chandran]. (It may soon be possible
to run hybrid particle-fluid simulations that will address this problem
directly. In addition, the observational
front, the failure to detect TeV $\gamma$-rays from some young 
supernova remnants has been interpreted as implying that many remnants
do not conform to the theory of shock acceleration and that
extra factors are required to account for the overall spectrum of Galactic
cosmic rays. 

A central theoretical 
problem with shock acceleration is that we do not have a generally
accepted understanding of particle injection. We do observe that
electrons appear to be injected into the acceleration process 
at a slower rate than ions. Undoubtedly the plasma physics is quite
complicated and Bernstein modes provide a plausible way to 
control this process [Dendy]. 

It appears that most of the compact sources that we observe, as well
as a good fraction of the type II extended radio sources, are energised 
by relativistic shocks. The standard theory of Fermi acceleration [Quenby] 
cannot apply here because the energy changes in a single shock 
passage are not small. Nonetheless the problem can be well-posed
if sufficiently stringent assumptions are made about
the transport. This problem has recently been solved by several groups
and a logarithmic slope of 2.23 derived for the
particle energy spectrum [Kirk]. It is not clear how relevant 
this is when the shock normal speed exceeds $c\cot\theta$ and it is
not possible to transform away the electric field. However it certainly 
represents an important advance and provides a basis for addressing the
next problem which concerns the effect of the back reaction on the
accelerated particles mediating the shock structure.  

It is no less important to understand how magnetic field is amplified
at a shock front. We fully expect that there will be adiabatic
compression at a non-relativistic shock so that 
the rms field shock strength
will increase by a factor $(\sin^2\theta+r^2\cos^2\theta)^{1/2}$, 
where $r$ is the compression ratio and $\theta$ is the angle between the 
magnetic field direction and the shock normal. 
($r=4$ for a non-relativistic
strong shock in ionised gas.) However, it has commonly been assumed,
for example in
studies of gamma ray bursts, that the post-shock field rapidly builds up to
a significant fraction of the equipartition value
[Birkinshaw]. This can orders
of magnitude larger than the compression value.  I do not think that there
is any good understanding of what to use in any of these environments.
Although there have been suggestions of post-shock turbulence in studies 
of supernova remnants, it has more commonly
been assumed that field amplification occurs around the generalised
contact discontinuity where the shock interstellar medium interact directly
with the ejecta from the supernova explosion. The fact that we find it so hard
to locate the bounding shock fronts around supernova remnants, whereas
we commonly observe the non-thermal synchrotron emission in the 
form of a diffuse, expanding ring, supports this view. Perhaps relativistic
shocks are different. Pulsar nebulae, containing ultrarelativistic, 
hydromagnetic shocks are the best laboratories that we have to 
study, {\it in situ} particle acceleration in an environment believed to be
similar to that in extragalactic, relativistic jets and gamma ray bursts.

The global evolution of the magnetic field is also highly relevant. There is 
increasing evidence for helical field structure in the observed jets
[Laing, Gabuzda], though 
it is quite unlikely that this field can be unidirectional. The total magnetic 
flux that would be necessary to make this happen is far larger than
could ever be confined in the vicinity of the central black hole.
Again, our understanding is very primitive.

The new images of X-ray jets presented and discussed
here [Harris, Wilson, Worrall, Sparks] are full of surprises.
Most encouragingly, they appear to be quite common and some patterns
are starting to emerge [Birkinshaw]. In addition, Pictor A is strikingly
well-collimated and it was particularly reasssuring to 
see the jets and lobes of Cygnus A revealed in this manner.

However, the interpetations were quite varied.
The case is well made that there is no simple common explanation
for the emission. Synchrotron radiation (requiring the accleration of 100
TeV electrons) [Harris], 
inverse Compton emission (of local synchrotron photons,
the ambient, external radiation field [Brunetti]
and the microwave background radiation[Tashino])
and, perhaps even in a few cases, bremsstrahlung 
or proton synchrotron radiation [Protheroe] can all 
occur under different conditions.
Most importantly, there is a working, synchrotron-inverse Compton model
of the integrated spectrum that makes a lot of sense in relation to what
we know about radio jets [Celotti].

In retrospect this should not be a surprise. After all, we directly 
observe jets on scales from milli- to Megaparsecs and over a wide range of
physical conditions when different
emission mechanisms can dominate. As the cooling times 
of the relativistic electrons
are generally quite short compared with the dynamical times, the X-rays
provide an invaluable probe of the conditions within these sources.
The effects of relativistic Doppler boosting, 
which we now take for granted, are also turning out to be quite subtle
[Celotti].
This is also not surprising because it is easy to highlight a dynamically
insignificant part of the source when it is directed towards us [Hardcastle].
These are early days in the study of X-ray jets.

We have recently had cause to question the assumption that
the radio emission from compact radio sources is due to the 
synchrotron process. The 
measured ``variability'' brightness temperatures can be nine orders of
magnitude above the standard inverse Compton limit. 
Although we are no longer intimidated by the large bulk Lorentz factors 
that are implied ($\Gamma>10^3$) - gamma ray bursts have released us from 
our inhibitions - the derived radiative efficiency is likely to be 
impossibly low. For this reason, there has been considerable attention 
paid to the influence of refractive scintillation on variability.
There have been some exquisite observational investigations 
[Dennett-Thorpe, Jauncey, Wagner] which have demonstrated that the 
observed variations are 
``sins of transmission''. However the source brightness temperatures
still have to be challengingly large ($>4\times10^{13}$~K 
corresponding to $\Gamma>30$ in the best studied
case and probably much larger in other
cases). The motivation for considering alternatives
to synchrotron emission remains and shock fronts, which we know can lead
to ring-like particle distribution functions that are far from equilibrium
together with high energy densities of plasma wave modes, 
are natural sites for non-linear
conversion to electromagnetic modes and maser emission [Bingham]. 
These possibilities deserve further study.
\subsection{What are Jets?}
It is a matter of considerable embarassment that we still cannot identify
the working substance in the jets.  This material is 
surely not immutable. Just as 
we are confident is the case in the Crab Nebula, the energy is 
probably changed
from an electromagnetic form to a pair plasma to an ion plasma and (at least
partially along the way into photons). We expect that this also happens
in jets. However we do not have a good understanding of where 
the transitions occur and, in particular, what we are observing using
EGRET, VLBI, MERLIN and the VLA! 

The reason why something like this 
must happen at small radius is that radiation drag is catastrophic 
unless there is an alternative carrier of momentum to pair plasmas.
The only two plausible candidates are protons [Protheroe] and Poynting flux
[Meier, Begelman]. 
The former has always seemed rather problematic because it is not at all clear
how to get the high energy protons in the first place (shocks only 
randomise the energy that is already present), without invoking an even greater
energy density in electromagnetic field. This field may also have to be
invoked to collimate the outflow. If electromagnetic field 
is present the electric fields are likely to be so strong that
relatively small inductive effects will
create a non-zero $\vec E\cdot\vec B$ which will break down the vacuum 
by creating copious electron-positron pairs, just as is
envisaged in a pulsar magnetosphere.  

The final conversion to an ionic plasma
seems equally inevitable because a jet must have an outer radius and it seems
very hard to prevent some measure of entrainment of the surrounding plasma 
along its length, even if it is protected by a cocoon of backflowing 
plasma [Laing, Simkin]. 

There are three more direct, observational arguments that have been advanced 
in favour of pair plasmas being present in the radio emission regions.
Firstly, the presence of circular polarisation now in over twenty 
sources like 3C279,
where the the Faraday rotation appears not to be too large [Wardle]
is very difficult to explain without appealing to a pair plasma. (It is 
difficult to rule out synchrotron emission completely as the cause
but the frequency-dependence is not as predicted.) The second argument is 
dynamical [Celotti, Harris]. The estimated jet powers are excessive if the 
relativistic electrons that have to be invoked to carry the emission carry 
protonic, as opposed to electronic, baggage. The third argument is quite
ingenious and comes from the analysis of narrow emission 
lines formed when jets impact interstellar clouds. The line ratios
are not as predicted by conventional ionising agents and may indicate
that relativistic particles are at least partly responsible. It may 
be possible to distinguish positrons and protons by following this 
approach. However, none of these
arguments is, as yet, watertight, though there is every reason to 
be optimistic that they will strengthen with more 
X-ray and, ultimately, $\gamma$-ray  observations. 

Of course if the 
presence of pair 
plasma can be demonstrated, this very much strengthens the argument 
that jets are powered by spinning black holes. 
\section{Lobes}
\subsection{What is a Lobe?}
The short answer to this question has been known since the early 1970s. 
A lobe is the repository for all the energy that is left over after the 
momentum of a jet has been exhausted pushing away the intergalactic medium.
It is the fact that the speed of the jet is much greater than
the speed of advance of the lobe that leads to this surplus.

We are now able to study lobe morphology in considerable detail
and able to address much more sophisticated questions. 
The key to understanding the overall energetics of radio
sources is to measure the pressures
within and around the lobes. This can be done using X-ray observations.
However, it appears that the total energy density in the 
lobes can significantly exceed the absolute minimum energy density 
required to account for the measured synchrotron radio emission; the difference
probably being made up with high energy electrons [Leahy, Blundell].
\subsection{Lobe Energetics} 
Radio sources do have a pronounced effect on the
overall appearance of many clusters and large ``holes'' can be seen
in the X-ray maps that correspond to the location 
of the radio lobes [Fabian]. The characteristic 
shapes of prominent radio lobes, like those associated with
Cygnus A, may be a straightforward consequence of the 
jet dynamics and reflect the overall stability of the 
jet [Eilek, Rudnick]. The sharply-defined boundaries that 
are seen are important because they set an empirical upper bound on the rate
at which relativistic and thermal plasma can mix [Owen, Rudnick,
Reynolds].
Unfortunately, many of these X-ray maps are turning out ot be 
very complex and, as we do not know the three dimensional distribution of 
the gas,
it is going to be quite hard, in practice to interpret the Sunyaev-Zel'dovich
maps with much precision. 

A recent development is the advent of hybrid codes that are not only
able to trace the hydrodynamical evolution of a jet, together 
with its associated lobe, but can also follow the passive transport
of magnetic flux and relativistic particles accelerated in shock fronts
[Tregellis, Jones, Marcowith]. Large scale magnetic field 
amplification and turbulence evolution 
can also occur [De Young]. Although the models will never correspond 
in detail to individual sources, there is some optimism that we will be 
able to understand the rules by which particle energisation and transport
and field amplification take place under relativistic conditions.
As is the case with jets, not all of the particle acceleration appears
to be associated with strong shock fronts.
Ongoing particle acceleration at sites other than strong shock fronts,
is often required because the radiative 
loss timescales can be shorter than the source lifetimes
and backflow times [Laing]. Magnetohydrodynamic
turbulence, perhaps associated with noise from the supersonic jet provides
a very plausible source of the long wavelength waves that can energise 
the particles in the lobes. Adiabatic compression can contribute
to the re-energisation under some circumstances [Ennslin]. 
\subsection{When did Lobes Form?}
The ability to trace the lobes to very low surface brightness is starting
to allow us to understand the ``archaeology'' of double radio sources. In
sources like M87, it is, in principle, possible to trace the history
of nuclear activity for the past Gyr [Begelman]. 
The large scale, low radio frequency
structure of this source is actually reminiscent of a ``mushroom cloud''
and could arise for essentially the same reason [Kaiser].
This is particularly interesting 
because, in this example, we know, from the mass of the central black hole
($\sim3\times10^9$~M$_\odot$), that the source was almost certainly 
once a quasar, although probably not recently. 
At the very least, the source behaviour has been quite variable
[Binney].  It is likely that the 
lobe will rise in the cluster gravitational field under buoyancy, until it
finds a static equilibrium state.

Spectral aging studies of radio lobes are becoming increasingly sophisticated
[Blundell, Rudnick]. Although particle and field injection 
is controlled by the hot spots, whose power declines with time, there
is much more going on. Again, the validity of equipartition [Leahy, Hardcastle]
and the possibility that the emitting plasma have a small filling
factor [Manolakou] is again under debate. At the very least, 
the dangers of interpreting inadequately resolved radio maps 
are now quite clear.
\section{What is the Environmental Impact?}
The Chandra observations of clusters
appear to corroborate the view that, in many rich clusters,
the central cooling times are short compared with the age of the universe. What
happens next remains controversial. The most natural consequence is that 
gas rains down upon the galactic nucleus. However, in several cases,
we are not seeing the evidence for cooler gas. There have been several 
suggestions as to how to prevent this from happening. The gas may be reheated
by intergalactic supernovae (although these are not observed) or low energy
cosmic rays. Alternatively, there may be a
large turbulent, magnetic or cosmic ray pressure. 

This last possibility seems particularly interesting and has
broader implications. The hot gas that we observe in clusters,
groups and elliptical galaxies has presumably been heated 
by passing through a shock front; there really is no alternative.
This is generally supposed to be an accretion shock located at
the cluster periphery. However it could be that there is also a dense
network of intergalactic shock waves and that most gas
is heated there.  These intergalactic shocks are quite likely 
to arise because
the sound speed in the ionised (Ly$\alpha$ cloud)
intergalactic medium is $\sim10$~km s$^{-1}$,
whereas the chaaracteristic random motion of field galaxies (including
our own) is several hundred km s$^{-1}$. The ionised gas that we observe
at high redshift seems to disappear and it is unlikely that it 
forms stars and galaxies with extreme efficiency. A far more plausible fate
is that it is rendered effectively invisible by heating to a million 
degrees. (Although it is possible that FUSE or XMM-Newton may find
most of the local intergalactic medium, we will probably have to wait until  
the next generation of X-ray telescopes is launched). The intergalactic
medium may be just as active and as interesting as the interstellar
medium has been shown to be.
Now if Galactic supernova remants are any guide, then the transmited partial
pressure in GeV ions by these shocks may be compable to the pressure of
thermal gas [Jones]. This will alter our estimates of the total 
static pressures
in clusters and, consequently, radio source energetics.  More interestingly
and importantly,
it is possible that these cosmic rays will inhibit inflow after cluster
gas starts to cool. 
\section{How Important are Extragalactic Radio Sources?}
Although the study of extragalactic radio sources retains its intrinsic
interest, it is (or should be)
attracting the attention of a much larger community. The most timely 
connection is to the study of gamma ray bursts. These are now commonly
(though still not securely) modeled as 
ultrarelativistic beamed outflows - jets.
Many of the principles that were debated and discussed twenty years ago
in the present context are being recapitulated in connection with 
gamma ray bursts
[Rees]. There are, however, some important differences. Firstly,  burst
powers (typically estimated to be $\sim10^{44}$~W sterad$^{-1}$) 
are many orders of magnitude
larger than the strongest quasar jets (up to 
$\sim10^{40}$~W sterad$^{-1}$).
Secondly the power is generally impulsive and so, although quasars
are increasingly regarded as episodic, there is usually a pre-existing
channel for them to propagate along. This difference is responsible 
for the observation of discrete afterglows associated with bursts,
which overlap in the case
of jets. Thirdly, the variability 
times observed in gamma ray bursts are very much shorter
than in their AGN counterparts and this implies that GRB 
sources are much more compact. This implies much larger initial
Thomson and pair production optical depths which, in turn, are 
responsible for the larger terminal Lorentz factors. However, the difference
appears to be only one of degree. 

Most tantalising of all,
though, are the reports of considerable iron line
emission [Rees]. If true, this is probably a 
big clue to the identity of the underlying sources. 
Otherwise, the similarities in and 
debates about the source models energy pathways and emission 
mechanisms are quite strong and there is a convergence between the two 
research fields which should be exploited. For example, now that 
attention has settled on jet models of gamma ray bursts, it probably 
becoems necessary to consider oblique relativistic shocks as the 
emitting elements. This will have a significant impact on the kinematics.
Conversely, if we are really observing jets with $\Gamma\sim300$
in bursts, then this should help us decide if really are
observing flows with comparable speeds in AGN.

Another overlap, surprisingly not represented at this meeting, is the study
of the Galactic superluminal sources into which some gamma ray bursters
may eventually evolve. So far, the best Galactic cases have only 
exhibited modest superluminal expansion speeds, though this could be a 
selection effect. Perhaps we have to explore other galaxies 
to find micro-3C273's. The advantage of studying such, sources, if they do
exist, is that they vary and cycle on timescales that are six to eight
orders of magnitude shorter than their quasar counterparts and obsevers
would get far more immediate gratification if they could observe such sources.
For example, we are awash with data on 
QPOs in the best studied case, GRS1915+105, whereas it
has proved very hard to identify corresponding phenomena despite  
thirty five years of quasar monitoring. The same may happen witr
relativistic jets.

A further possible point of contact that was discussed [Watson], is the
origin of ultra high energy cosmic rays - protons, presumably, with
Lorentz factors up to $\sim3\times10^{11}$ and energies
of 50~J, comparable with that of the shot with which I was
caught out in the cricket match. These particles have lifetimes 
against photo-pion production in teh microwave background of
only a hundred million years. This tells immediately that
their luminosity density is quite large, in fact comparable with the 
average cosmological luminosity density of $\sim$~GeV cosmic rays.

There are two
classes of explanation -- that they are left over from the decay of
much higher energy particles, 
almost certainly formed cosmologically, or that they are 
accelerated from lower energy, just like all the other 
cosmic rays. The former explanation is surely the more exciting. However,
there appears to be a serious objection to it. It is very hard to avoid
producing gamma rays in far greater profusion than protons and these
can be distinguished statistically by their zenith angle distribution.
They are not seen. Furthermore, the same electromagnetic
channel is likely make a much larger TeV gamma background than is measured
(Coppi, private communication). For this reason interest has returned
to ``conventional'' astrophysical models. Most of
these models require potential differences $\Phi>3\times10^{20}$~V 
in order to accelerate the very highest energy particles and have 
associated powers $L>\Phi^2/Z_0\sim3\times10^{38}$~W, 
where $Z_0\sim377\Omega$, is the impedance of 
free space. One of the few source candidates we have that satisfies
the proximity and power requirements is the powerful extended radio galaxies.
(Black holes and compact jets are capable of 
accelerating ultra high energy cosmic rays 
but have too large an ambient radiation 
energy density to allow them to escape.)

My final example of the new relevance of extragalactic radio sources 
brings me back 
to the start of this summary. Outflows
from active galactic nuclei may also provide a feedback
that strongly influences, limits or even initiates 
galaxy formation. Perhaps relativistic jets
can deposit enough energy or
momentum in the surrounding gas to prevent disk formation in 
elliptical galaxies and in cooling flows
in the centres of clusters. Now radio sources themselves may not
be quite up to this job. In fact it has been shown that the gas surrounding
the radio lobes in at least one clusters is actually cool not 
hot [Fabian, Reynolds]. However,
hydromagnetic disk winds and, especially broad absorption 
line outflows could be more powerful and have a major impact. 
After all the nuclei of most 
powerful radio sources and central dominant galaxies 
in clusters almost surely passed through one or more 
quasar phases in the past and we expect that most of the energy
that was released will be during a hyperactive, quasar phase. 

This general idea has been rather controversial,
mainly because of its subversive implications for the current standard
model of galaxy formation. However, it is still worth exploring
with numerical simulation. Perhaps
the real ``purpose'' of outflows from accreting sources is to limit
their gas supply to a a manageable rate. 

\section{Peter Scheuer 19??-2001}
While completing this summary, I learned of the passing of Peter Scheuer.
Through his own major research contributions, his gentle 
yet incisive questioning and his selfless encouragement of others, Peter had
a lasting influence on every aspect of this meeting and on most
of the contributors. This article is dedicated to his memory. 
\section*{Acknowledgements}
I am indebted to the National Science Foundation and NASA for support under
grants AST 99-00866 and 5-2837 respectively. I thank the organisers for 
arranging an enjoyable workshop and their gracious hospitality.

\end{document}